\documentclass[10pt,conference]{IEEEtran}
\IEEEoverridecommandlockouts

\usepackage{cite}
\usepackage{amsmath,amssymb,amsfonts}
\usepackage{algorithmic}
\usepackage{graphicx}
\usepackage{textcomp}
\usepackage[dvipsnames]{xcolor}
\usepackage{enumitem}
\usepackage{blindtext}
\usepackage{collectbox}
\usepackage{xspace}
\usepackage{csquotes}
\usepackage[pdftex]{hyperref}
\usepackage{tcolorbox}
\usepackage{array}
\usepackage{booktabs}
\usepackage{subcaption}

\def\BibTeX{{\rm B\kern-.05em{\sc i\kern-.025em b}\kern-.08em
    T\kern-.1667em\lower.7ex\hbox{E}\kern-.125emX}}

\newboolean{showcomments} 
\setboolean{showcomments}{true}
\ifthenelse{\boolean{showcomments}}
{\newcommand{\nb}[2]{
		\fbox{\bfseries\sffamily\scriptsize#1}
		{\sf\small$\blacktriangleright$\textit{\textcolor{red}{#2}}$\blacktriangleleft$}
	}
}
{\newcommand{\nb}[2]{}}

\definecolor{my-blue}{RGB}{213,226,250}
\definecolor{azulbb}{RGB}{204, 230, 255}

\begin{document}

\title{Hearing the voice of experts: Unveiling Stack Exchange communities' knowledge of test smells}

\author{\IEEEauthorblockN{Luana Martins}
\IEEEauthorblockA{\textit{Institute of Computing} \\
\textit{Federal University of Bahia (UFBA)}\\
Salvador, Brazil \\
martins.luana@ufba.br}
\\
\IEEEauthorblockN{Joselito Mota Junior}
\IEEEauthorblockA{\textit{Institute of Computing} \\
\textit{Federal University of Bahia (UFBA)}\\
Salvador, Brazil \\
joselito.mota@ufba.br}
\and
\IEEEauthorblockN{Denivan Campos}
\IEEEauthorblockA{\textit{Institute of Computing} \\
\textit{Federal University of Bahia (UFBA)}\\
Salvador, Brazil \\
denivan.campos@ufba.br}
\\
\IEEEauthorblockN{Heitor Costa}
\IEEEauthorblockA{\textit{Department of Computer Science} \\
\textit{Federal University of Lavras (UFLA)}\\
Lavras, Brazil \\
heitor@ufla.br}
\and
\IEEEauthorblockN{Railana Santana}
\IEEEauthorblockA{\textit{Institute of Computing} \\
\textit{Federal University of Bahia (UFBA)}\\
Salvador, Brazil \\
railana.santana@ufba.br}
\\
\IEEEauthorblockN{Ivan Machado}
\IEEEauthorblockA{\textit{Institute of Computing} \\
\textit{Federal University of Bahia (UFBA)}\\
Salvador, Brazil \\
ivan.machado@ufba.br}
}

\maketitle

\begin{abstract}

Refactorings are transformations to improve the code design without changing overall functionality and observable behavior. During the refactoring process of smelly test code, practitioners may struggle to identify refactoring candidates and define and apply corrective strategies. This paper reports on an empirical study aimed at understanding how test smells and test refactorings are discussed on the Stack Exchange network. Developers commonly count on Stack Exchange to pick the brains of the wise, i.e., to `look up' how others are completing similar tasks. Therefore, in light of data from the Stack Exchange discussion topics, we could examine how developers understand and perceive test smells, the corrective actions they take to handle them, and the challenges they face when refactoring test code aiming to fix test smells. We observed that developers are interested in others' perceptions and hands-on experience handling test code issues. Besides, there is a clear indication that developers often ask whether test smells or anti-patterns are either good or bad testing practices than code-based refactoring recommendations. 

\end{abstract}


\section{Introduction}\label{sec:introduction}

Software refactoring consists of changing the structure of the source code without compromising its overall functionality \cite{alizadeh2018interactive} and observable behavior \cite{fowler2018refactoring}. We use software refactoring to enforce better design and coding practices through small code transformations \cite{Santana2021_MultiMethod}. In addition, software refactoring aims to facilitate readability and maintainability \cite{ReadaAndMaint}, especially in large code bases where multiple developers engage without a detailed view of the whole system \cite{christopoulou2012automated}.

Typically, the refactoring process takes place in the key phases \cite{kataoka2002quantitative, katic2009towards}: \textit{(i)} identification of candidates for refactoring; \textit{(ii)} determination of the appropriate refactoring technique; \textit{(iii)} application of refactoring; and \textit{(iv)} validation of the refactoring effect. Identifying a refactoring candidate requires an in-depth understanding of various parts of the system and knowledge of best practices \cite{oliveira2020collaborative}. Detecting problems in the code can be time-consuming and labor-intensive and often requires automated tool support to be effective in practice \cite{paiva2017evaluation}. Besides, deciding which refactoring technique best applies in each situation can be challenging and complex \cite{tempero2017barriers}.
Such reasons lead software developers to understand how the software community has dealt with anti-patterns and their refactoring processes \cite{Peruma2022_HowDoIRefactorThis}.

The Software Engineering research community has dedicated efforts to understanding anti-patterns and proposing solutions to assist developers in refactoring the code to cope with design issues \cite{palomba2014they, Spadini2020_SeverityThresholds, hozano2018you, Garousi2018_SurveySmells}. However, the evolution of programming languages and frameworks creates new anti-patterns that require novel refactorings to fix them, making it challenging for the community to keep up with the actual problems developers face in practice \cite{garousi2016challenges}. For this reason, developers often post questions on Q\&A platforms seeking help to solve a problem with their code. In a recent study, Peruma et al. \cite{Peruma2022_HowDoIRefactorThis} conducted quantitative and qualitative experiments to understand how developers discuss refactoring in a collaborative online discussion forum. The authors aimed to unveil the most explored and discussed topics concerning software refactoring. 

From a broader perspective, refactoring is just as important for automated test code as production code as it supports the identification of issues caused by production code changes \cite{karac2018we, dybaa2008empirical}. Testing verifies whether the software functionality and observable behavior are kept the same after code refactorings \cite{alomar2021preserving, Guerra2007_Refactoring_Safely}. The test result should remain unchanged before and after the refactorings in the production code. However, the test code development is prone to human errors, harming the test code's ability to detect defects. Therefore, to effectively diagnose problems in production code, Beck et al. \cite{beck2003test} stated the coding of tests should follow good design principles. 

Inspired by those arguments, van Deursen et al. \cite{Deursen2001_TestRefactoring} defined test smells to indicate badly designed tests \cite{Deursen2001_TestRefactoring}. The presence of test smells can be interpreted as a symptom of poor software quality, harming the testing and maintenance activities \cite{greiler2013automated,Peruma2019_DistributionAndroid, Spadini2020_SeverityThresholds, kochhar2019practitioners}. Garousi et al. \cite{Garousi2018_SurveySmells} proposed a catalog of test smells and a summary of existing techniques and tools resulting from a multivocal literature review. That catalog and summary are a good initial step towards advancing the field, but there is still a lack of understanding of which refactoring to apply and how to apply them, in practice, to fix test smells. 

By analyzing the literature on test smells \cite{Garousi2018_SurveySmells, Spadini2020_SeverityThresholds, kim2020empirical}, we may observe a few pieces of evidence of the challenges developers face during the automated test code development and maintenance and the commonly discussed issues, mainly unit testing. Furthermore, little is known about commonly used corrective strategies and strategies that ensure the preservation of test behavior after test code refactoring. These are important gaps to bridge.

This paper reports on the results of an empirical study aimed at understanding how the developers discuss test smells and test refactorings in the Stack Exchange, the leading software development collaboration network. By analyzing data from Stack Exchange, we could contribute with a synthesis of the community discussions about test smells, mainly regarding the key challenges, test design issues, test code refactoring, and post-refactoring test behavior.

\section{Research methodology}\label{sec:researchmethodolog}

This study addressed the following Research Questions (RQ):

\begin{enumerate}[label=\bf RQ\textsubscript{\arabic*},left=.5cm]

    \item  \textbf{What challenges do developers report for handling problems in the test code?} This RQ studies developers' main difficulties in refactoring test code by grouping the questions developers ask into \textit{why-how-what} categories.

    \item \textbf{What test smells do developers most actively discuss?} This RQ studies which test smells developers consider relevant to refactor and classifies them following Garousi et al.'s catalog \cite{Garousi2018_SurveySmells}.

    \item \textbf{What preventive and corrective actions do developers suggest to handle test smells in the test code?} This RQ leverages the actions developers suggest to prevent test smell insertion and the refactoring operations from fixing test smells.
    
    \item \textbf{Do developers discuss how to keep the test behavior after test code refactorings?} This RQ investigates whether and how developers care about test behavior during refactoring.

\end{enumerate}

\begin{figure*}[tb!]
\begin{center}
{\includegraphics[width=1\linewidth]{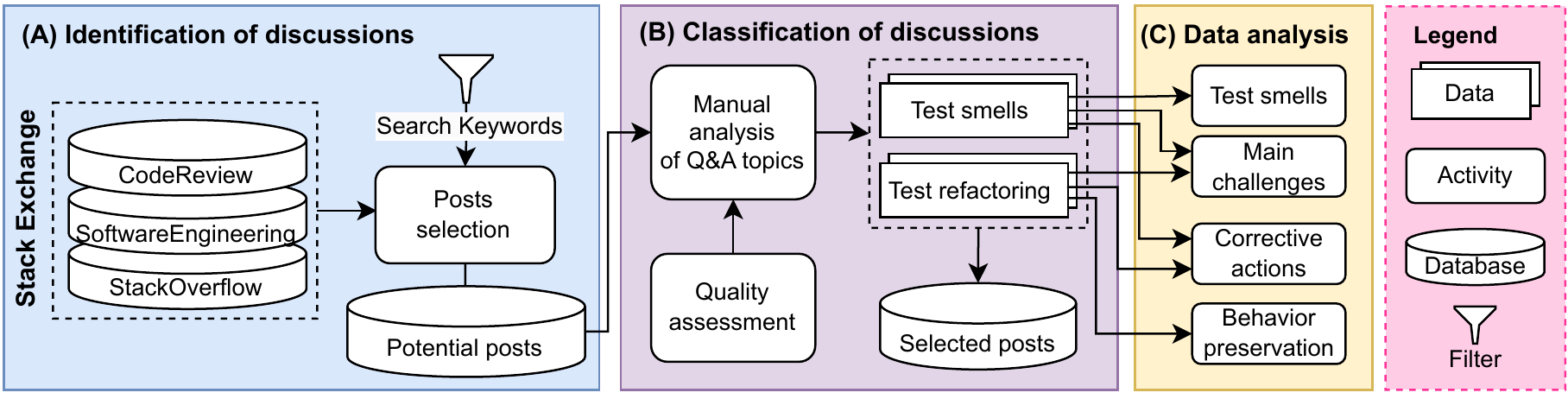}}
\caption{Study Design}
\label{fig:design}
\vspace{-0.5cm}
\end{center}
\end{figure*}

Fig. \ref{fig:design} shows the study design, which encompasses three main steps: (A) Identification of 
discussions, 
~(B) Classification of discussions,
~and  (C) Data analysis.

\vspace{1em}

\subsection{Identification of 
discussions}
\label{subsection:identification-of-discussions}

Developers’ competence consists of a frequent lifelong quest for knowledge, and educational networks are excellent environments for spreading knowledge \cite{StackExchange}. According to Tahir et al. \cite{Tahir2020_CodeSmells}, developers often gather in Q\&A online forums to `look up' how others complete similar tasks and cope with recurring issues, including how to get rid of anti-patterns. For example, \textit{Stack Exchange} is a collaborative discussion forum network offering insightful resources on development issues \cite{Tahir2020_CodeSmells}. Posnett et al. \cite{StackExchange} mentioned participating in and sustaining learning communities is a durable and valuable aspect of professional life.

In this study, we used Internet Archive (IA)\footnote{Available at \url{https://archive.org/download/stackexchange}} to retrieve discussions on test code problems developers report and the corrective actions they suggest. IA keeps data dumps of discussions on the Stack Exchange network over time. We selected the following Stack Exchange sites:

\begin{itemize}
    \item \textbf{Code Review\footnote{Available at \url{https://codereview.stackexchange.com/}}:} a site for peer programmer code reviews. It allows programmers to ask questions about specific code snippets and receive feedback from others;

    \item \textbf{Software Engineering\footnote{Available at \url{https://softwareengineering.stackexchange.com/}}:} a site for developers and scholars interested in asking general questions on the systems development life cycle \cite{Tahir2020_CodeSmells}. The site is adequate for opinion-based questions;

    \item \textbf{Stack Overflow\footnote{Available at \url{https://stackoverflow.com/}}:} a site covering technical and general discussions on problems unique to software development \cite{Openja2020_ReleaseEngineering, bhasin2021student}. The discussions commonly focus on specific programming problems or tools.
\end{itemize}

\begin{figure}[th!]
{\includegraphics[width=\linewidth]{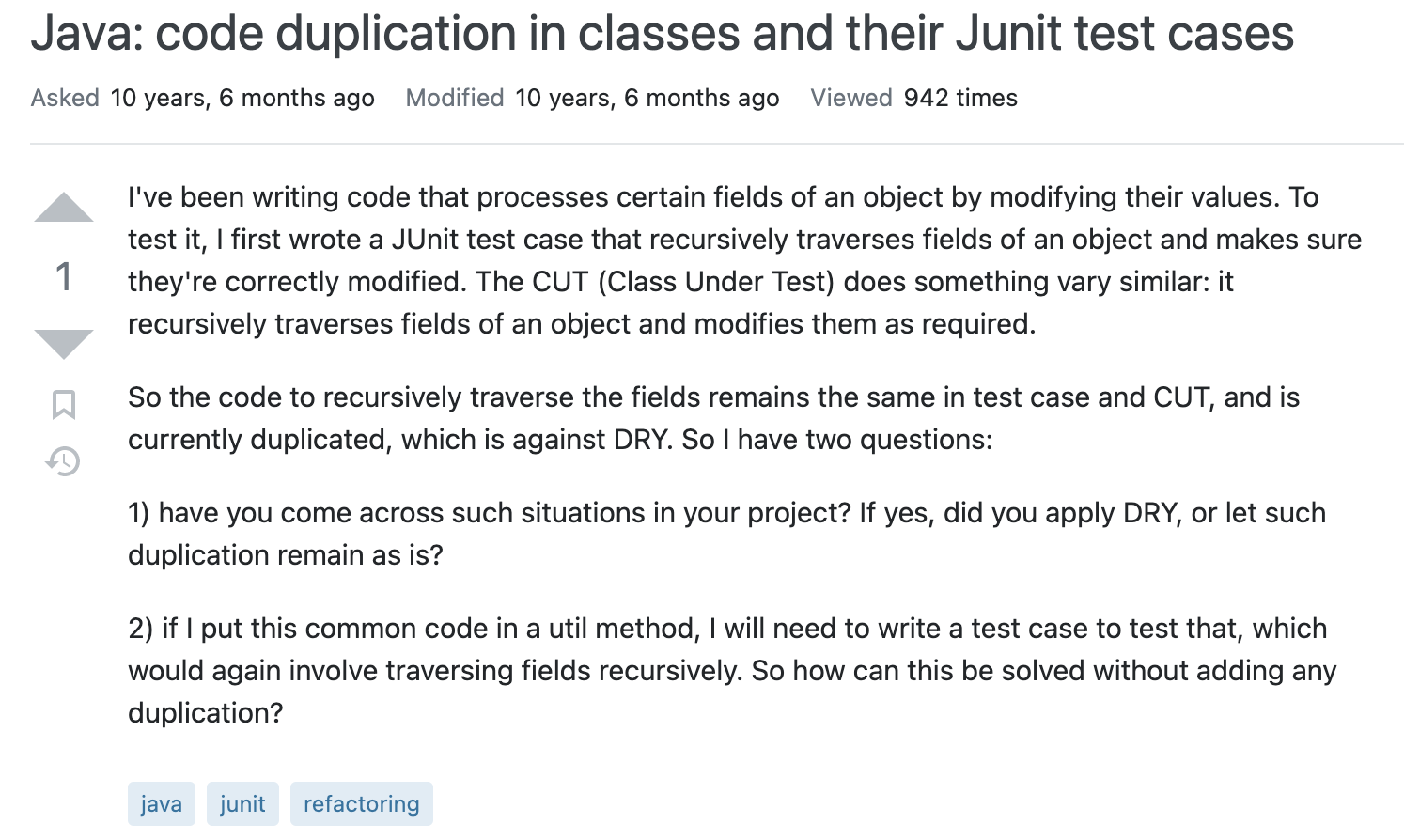}}
\caption{Sample question post 
extracted from StackOverflow.}
\label{fig:post-example}
\end{figure}

\begin{figure}[th!]
{\includegraphics[width=\linewidth]{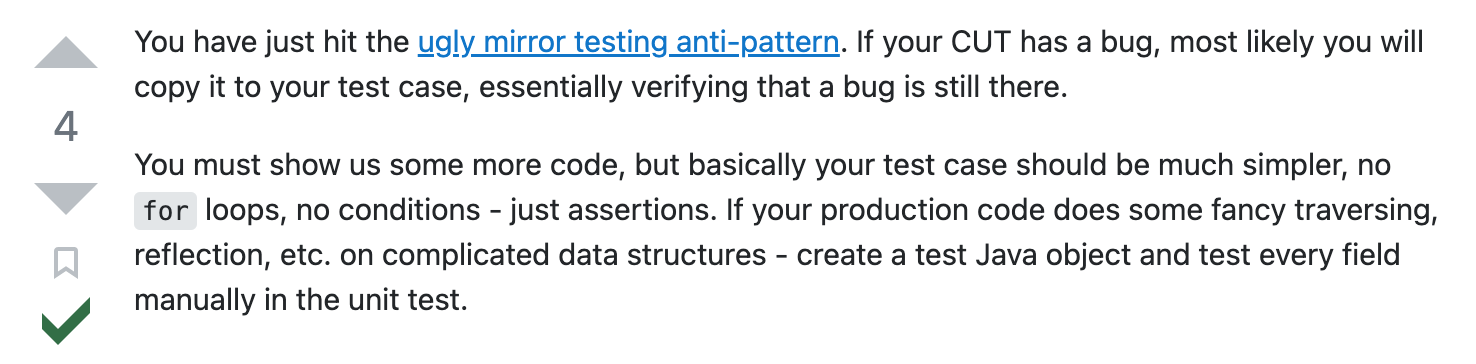}}
\caption{Accepted/useful answer from a StackOverflow post.}
\label{fig:answer-example}
\end{figure}

The discussions on the Stack Exchange sites associate one question post with one or more answer posts given by different users. A question post consists of a title, body, and tags. Fig. \ref{fig:post-example} shows a sample StackOverflow question post \cite{examplePost}. It comes with three tags: \texttt{Java}, \texttt{JUnit}, and \texttt{refactoring}. In the post, the author asks if the practice {\color{black}she/he} adopted in the \texttt{JUnit} test case consists of duplication and how {\color{black}she/he} writes the test without duplication. Fig. \ref{fig:answer-example} shows a response to the question post from Fig. \ref{fig:post-example}. It reveals that the practice of the question refers to a test anti-pattern called \texttt{Ugly Mirror}. As a potential solution, the test code simplification (avoiding conditional structures and using assertions instead) and the test object creation to run the tests manually for complex data structures. The post's author rated the answer as accepted, and the answer got four votes.

\begin{table*}[tb]
\scriptsize
\centering
\def \arraystretch{1.2}
\caption{Groups of tags composing the search string.}
\label{table:search-string}
\begin{tabular}{clm{3.2cm}m{9.5cm}m{1.0cm}} \toprule
\textbf{ID} & \textbf{Group} & \textbf{Description} & \textbf{Tags} & \textbf{Criteria}\\
\midrule

1 & TEST & Words related to testing, test type, or test structure & `unit-testing,' `testing,' `test-scenarios,' `unit-test-data,' `junit,' `junit4,' `junit5,' `automated-tests,'  `test-automation,' `tests,' `testcase,' `assertions,' `assert,' `assertion,' `annotations' & Inclusion\\

2 & DESIGN & Words related to programming practices and design & `code-smell,' `code-smells,' `anti-patterns,' `programming-practices,' `naming,' `naming-conventions,' `naming-standards,' `coding-standards,' `coding-style,' `code-formatting,' `format,' `formatting,' `bad-code,' `technical-debt' & Inclusion\\

3 & REFACTORING & Words related to refactoring & `refactoring,' `test-refactoring,' `automated-refactoring' & Inclusion\\

4 & LANGUAGE & Words related to programming languages & `c++,' `c\#,' `javascript,' `vb6', `python,' `python-3', `go,' `c,' `.net,' `php,' `sql,' `ruby' & Exclusion\\
\bottomrule
\end{tabular}
\end{table*}

We defined and applied a search string on the tags in each question post to select the discussions. It aimed at  filtering out the discussion content based on selected tags (Fig. \ref{fig:design} - \textit{Posts selection}). Table \ref{table:search-string} shows four groups of tags composing our search string. The \textit{TEST} group filters the discussions about test codes by specifying the types of tests, testing frameworks, and language constructs used in test codes. The \textit{DESIGN} group filters the discussions containing smell-related problems concerning test code \cite{Garousi2018_SurveySmells}. The \textit{REFACTORING} group filters topics with discussions on how to refactor test code. Considering that the tagged topics rarely use the words of the \textit{REFACTORING} group, we used a logical disjunction with the \textit{REFACTORING} and \textit{DESIGN} groups to create a less strict filter. Similarly, many topics are untagged with programming languages. Therefore, the group \textit{LANGUAGE} removes those topics tagged with programming languages other than \texttt{Java}.

Although some studies automatically classify thousands of Stack Exchange topics \cite{Peruma2022_HowDoIRefactorThis, Tahir2020_CodeSmells, Choi2015_CodeClones}, we fine-tuned our search string to reduce the number of non-relevant topics on test smells and test refactorings \cite{Savio2022, Savio2023}. We applied the search string ``\textbf{\textit{TEST AND (DESIGN OR REFACTORING) NAND LANGUAGE}}'' on the data dumps of three StackExange sites from September 15th, 2008 to December 6th, 2022. As a result, we retrieved 303 potential posts.

\begin{table}[tb]
\scriptsize
\centering
\setlength{\tabcolsep}{0.35em}
\def \arraystretch{1.3}
\caption{Number or retrieved discussions.}
\label{table:retrieved-discussions}
\begin{tabular}{lrrr} \toprule
\textbf{Site} & \textbf{\# Total posts} & \textbf{\# Potential posts} & \textbf{\# Selected posts} \\ \midrule
SoftwareEngineering & 237,548 & 153 & 40 \\
CodeReview & 196,301 & 2 & 2 \\
StackOverflow & +8million & 158 & 59 \\ \bottomrule
\end{tabular}
\end{table}

\subsection{Classification of discussions}
\label{subsection:classification-discussions}

We manually analyzed the discussions to select the ones related to  test smells and the refactorings to solve them (Fig. \ref{fig:design} - \textit{Manual analysis}). To align the analysis criteria, three coders performed a peer analysis on the CodeReview and SoftwareEngineering question topics. The coders read 155 question posts and applied the following inclusion (\textit{IC}) and exclusion criteria (\textit{EC}): \textit{($IC_1$)} question topics describing a problem in the test code related to bad design or implementation choices, \textit{($EC_1$)} question topics about the need for testing and how to create test code, and \textit{($EC_2)$} question topics without answers. We calculated the Kappa statistics \cite{Cohen1960} to assess the reliability of the manual classification (Fig. \ref{fig:design} - \textit{Quality assessment}) and reached a substantial agreement level of 0.61.\looseness=-1

After, independent coders analyzed the StackOverflow question posts and accepted 59 pots (Fig. \ref{fig:design} - \textit{Selected posts}, Table \ref{table:retrieved-discussions} - \textit{\#Selected posts}). Next, we analyzed 101 selected posts to extract the data items listed in Table \ref{table:data-items}. The data extraction followed the same steps described in this section. We performed the data extraction on the CodeReview and SoftwareEngineering in peers. The coders discussed divergences in data extraction in daily meetings and performed individual data extraction on StackOverflow. 

\begin{table}[tb]
\scriptsize
\centering
\caption{Data items extracted from discussions.}
\def \arraystretch{1.3}
\label{table:data-items}
\begin{tabular}{cm{2.3cm}m{4.1cm}c} \toprule
\textbf{\#} & \textbf{Data Item} & \textbf{Description} & \textbf{RQ}  \\ \midrule
D1 & Challenges for refactoring test smells & Challenges that developers face for refactoring the test code to fix test smells & RQ\textsubscript{1} \\ 
D2 & Description of test smells & Descriptions of test smells by developers based on their understanding & RQ\textsubscript{2} \\
D3 & Cause of test smells & Causes that lead to test smells & RQ\textsubscript{2}  \\
D4 & Actions for preventing test smells and refactoring the test code & Actions suggested by developers to prevent the insertion of test smells and refactor the test code to fix test smells & RQ\textsubscript{3}  \\
D5 & Tools for refactoring test smells & Tools and sources that support developers fixing test smells & RQ\textsubscript{3}  \\
D6 & Strategies to keep the test code behavior & Strategies to verify whether the test code behavior is kept the same after the test code refactorings & RQ\textsubscript{4}\\
\bottomrule
\end{tabular}
\end{table}

\subsection{Data Analysis}
\label{subsection:data-analysis}

To answer RQ$_1$, we analyzed the question topics in two steps. First, we applied a Thematic Content Analysis (TCA) \cite{braun2006using} to classify the types of questions asked by developers into \textit{why-how-what} questions (Golden-circle theory) \cite{Openja2020_ReleaseEngineering}: (1) \textit{why} is a type of question that seeks to understand the reason or the cause of a problem, (2) \textit{how} is the type of question that seeks approaches or better ways to achieve a result, and (3) \textit{what} is the type of question to get the information related to the problem. Then, we performed an open coding \cite{Stol2016GroundedTheory} to interpret the question topics and extract the main challenges developers face when applying refactorings to fix problems in the test code.   

To answer RQ$_2$, we applied a TCA to classify the discussions into (1) \textit{specific discussion} about a specific test smell and (2) \textit{general discussion} that does not explicitly ask about a test smell. After, we interpreted the description of the test smell and classified it according to Garousi et al.'s catalog \cite{Garousi2018_SurveySmells}. 

To answer RQ$_3$, we applied a TCA in the top answers (top-rated answer, accepted answer, or unique answer) to classify their actions into \cite{Tahir2020_CodeSmells}: (1) \textit{Fix} to recommend test code refactorings for fixing problems in the test code, (2) \textit{Capture} to explain the test problems but does not recommend test code refactoring to solve the problems, (3) \textit{Ignore} to recommend ignoring taking any action to fix the test code, and (4) \textit{Explain} why something is considered a test problem. In addition, we extracted the tools and applied open coding to list the test code refactorings suggested in the answers. 

To answer RQ$_4$, we analyzed whether the developers asked for strategies to verify the test code behavior after performing test code refactorings. We listed the strategies and tools suggested in the answers. Data is publicly available in an online open data repository \cite{paginaweb}.

\section{Results}\label{sec:results}

\subsection{Discussions characterization}

\begin{figure}[t]
\begin{center}
\begin{subfigure}{0.5\textwidth}
\includegraphics[width=\linewidth]{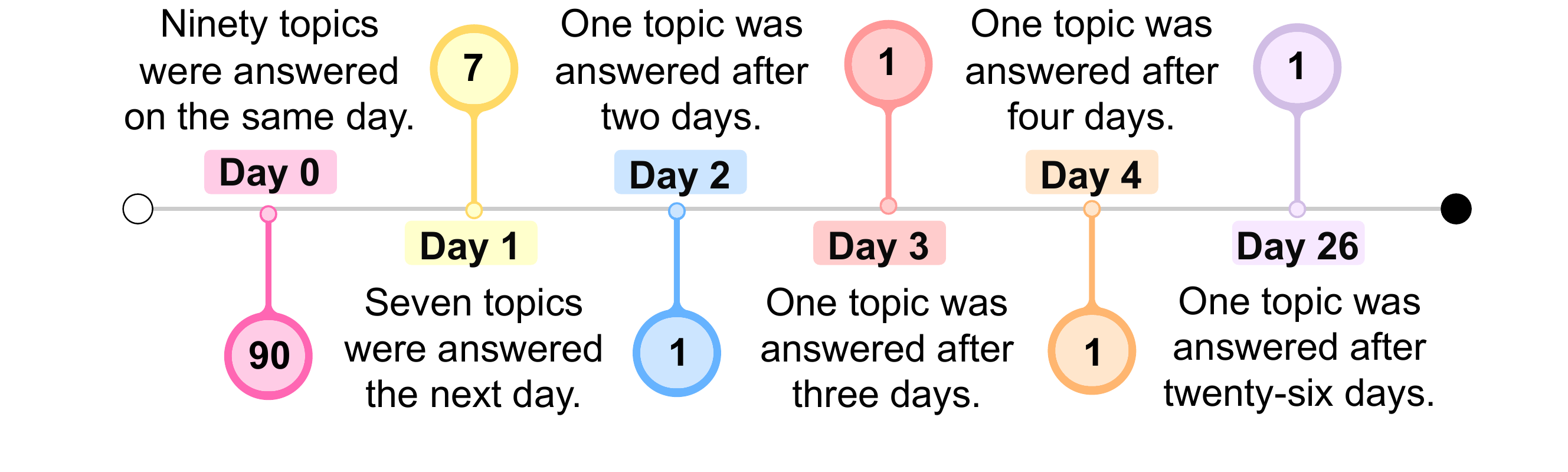} 
\caption{Time between asking a question and receiving a first answer.}
\label{fig:partA}
\end{subfigure}
\begin{subfigure}{0.5\textwidth}
\includegraphics[width=\linewidth]{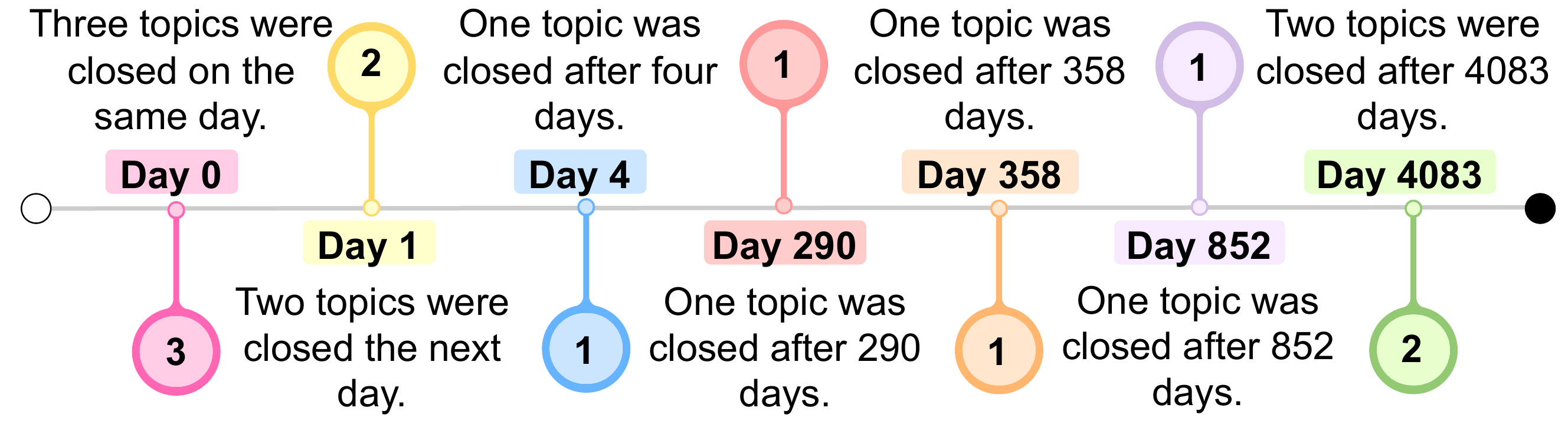}
\caption{Time between asking a question and closing the discussion.}
\label{fig:partB}
\end{subfigure}
\caption{Timeline of answer time and closing time of discussion topics in days}
\label{fig:responseandclosingtime}
\end{center}
\end{figure}

We characterized 101 discussion topics on test smells and their solutions found on the Stack Exchange network. 
There is nearly no repetition of users asking questions. For those users who asked more than one question, the author rephrased the question in a more specific way or on a different site. We collected 111 top answers, of which we accepted 61 answers, and 38 answers were top-rated. We found another 12 answers, and although not accepted, we deemed relevant for our investigation. Most users answered only one question, and only seven answered more than one. It is worth noting four out of these seven users are top-ten users when considering users' reputation score, number of questions, and number of answers. Next, we collected the time between asking a question and receiving the first answer (Fig. \ref{fig:partA}) and the time to close the discussion topic (Fig. \ref{fig:partB}). We observed that 90 out of 101 questions received an answer on the same day, and seven received an answer the next day. The remainder received an answer from the second to the twenty-sixth day. Regarding the time to close, only three topics were closed the same day they were posted, and the other 11 were closed within a year. Curiously, some took 4,083 days to close, and many never closed.

\subsection{Trends and Challenges (RQ1) }
\label{sec:results-test-challenges}

\begin{figure}[th]
\begin{center}
\vspace{-0.5cm}
{\includegraphics[width=0.9\linewidth]{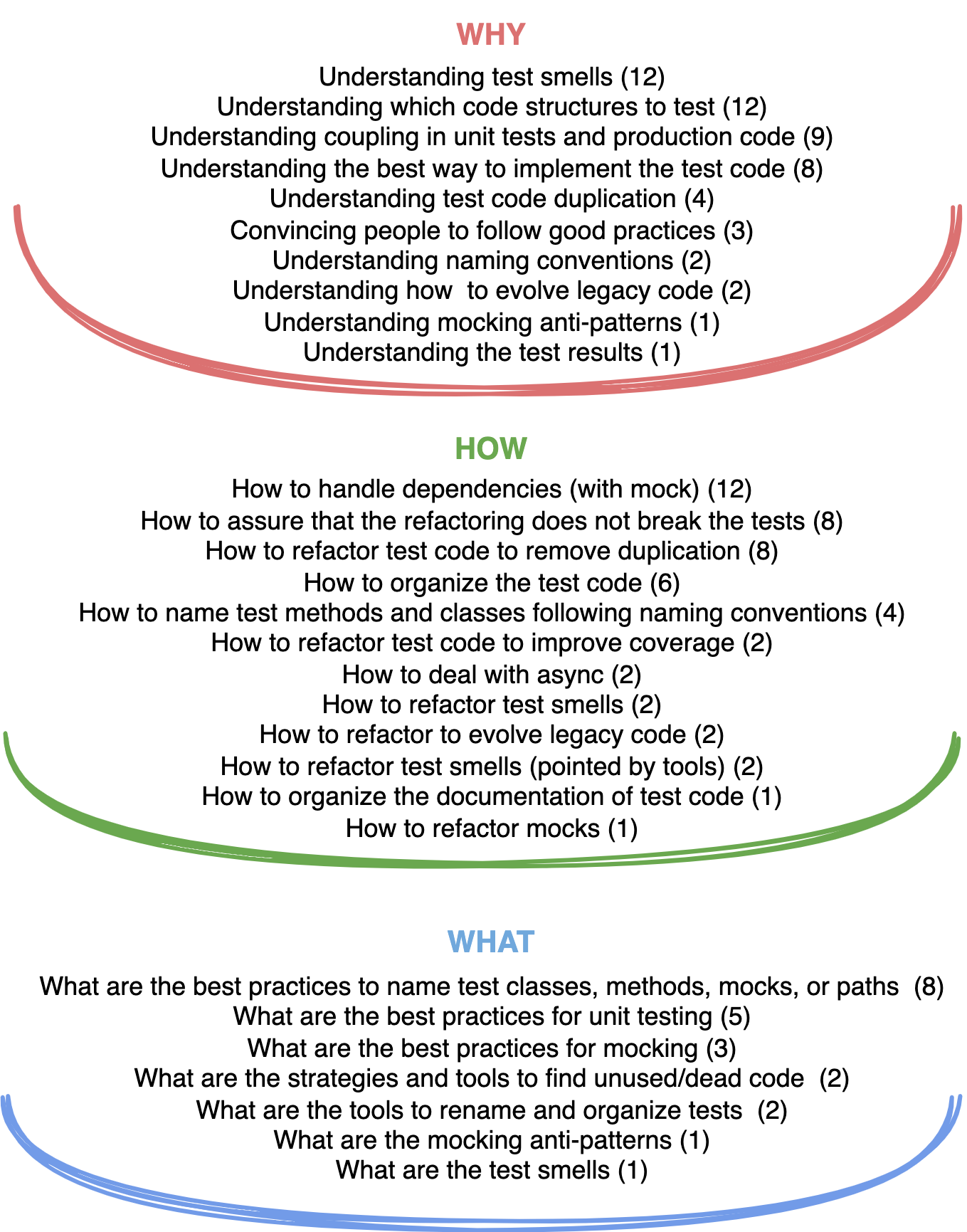}}
\caption{Classification of types of questions and challenges}
\label{fig:challenges}
\end{center}
\end{figure}

In RQ$_1$, we aimed to understand the trends and challenges around developers’ discussions on test code refactoring concepts and activities. We classified the discussions into 30 categories representing the questions asked by developers while evolving the test code. Fig. \ref{fig:challenges} presents the number of discussions in each category and their relationship with the \textit{why-how-what} questions.

In the \textit{why} questions, we classified the discussions into ten categories to understand the test code problems and bring insights into the decision-making of whether to refactor the test code for fixing such problems. The \textit{Understanding which code structures to test}, \textit{Understanding coupling in unit tests and production code}, and \textit{Understanding how to evolve legacy code} categories refer to problems with origin in the production code. The \textit{Convincing people to follow good practices} category relates to problems faced by newcomers joining a team that does not follow good practices for testing the code due to organizational decisions or team conventions. The \textit{Understanding test smells}, \textit{Understanding the best way to implement the test code}, \textit{Understanding test code duplication}, \textit{Understanding naming conventions}, \textit{Understanding mocking anti-patterns}, and \textit{Understanding the test results} categories refer to problems related to the comprehension of the testing frameworks constructs, the importance of following good practices and conventions for developing test code, and the problems caused by the wrong usage of such constructs, practices, and conventions.

We could sort most topics into \textit{Understanding which code structures to test} and \textit{Understanding of test smells} categories. In the former, the discussions usually present a code excerpt asking whether the developers should test a structure of the production code (e.g., overloaded/private/public methods). In the latter, the discussions often refer to whether a particular test code is either smelly or not and why someone should be concerned about the impacts of test smells.

For example, we classified the following question in the \textit{Understanding test smells} category. That category shows a barrier developers face in learning specific constructs for creating test cases. Although they perceive the code as smelly, they need to understand the pros/cons of improving the test code.

\vspace{.3em}
\begin{displayquote}
\footnotesize
\textsf{(...) I always tell myself as long as the ``real'' code is ``good,'' that's all that matters. Plus, unit testing usually requires various ``smelly hacks'' like stubbing functions.
How concerned should I be over poorly designed (``smelly'') unit tests? \cite{se_smellsMatter}}
\end{displayquote}
\vspace{.3em}

In the \textit{how} questions, we classified the discussions into twelve categories. These encompass strategies and techniques to handle problems reported in the \textit{why} questions through other developers' experience. The \textit{How to handle dependencies (with mock)} and \textit{How to refactor mocks} categories deal with the test code evolution to use mocks and stubs for testing external dependencies with APIs, databases, web services, and files. The \textit{How to refactor test code to remove duplication} category discusses how to handle duplication of setup methods, annotations, and test cases covering overloaded production methods. The \textit{How to deal with async} category discusses testing framework constructs to avoid non-deterministic tests given the asynchronicity of the production code. The \textit{How to refactor test code to improve coverage} and \textit{How to refactor to evolve legacy code} categories involve understanding the production code to co-evolve the production and test codes, improving test coverage. The \textit{How to organize the test code}, \textit{How to name test methods and classes following naming conventions}, and \textit{How to organize the documentation of test code} categories seek to standardize the naming, documentation, and organization of test codes. The \textit{How to refactor test smells} and \textit{How to refactor test smells (pointed by tools)} categories discuss strategies to fix test smells identified by developers or tools (e.g., SonarQube). Lastly, the \textit{How to assure that the refactoring does not break the tests} category presents discussions on techniques to verify whether the behavior of the test code keeps the same after test code refactorings.

For example, we classified the following question 
into two categories \textit{Understanding which code structures to test} and \textit{How to handle dependencies (with mock)}. 
The question shows an example of a helper method containing some external dependency. The first category aims to understand whether the developer should test the helper method, and the second asks for strategies to implement the test code handling the dependencies of the helper method. 

\vspace{.3em}
\begin{displayquote}
\footnotesize
    \textsf{So I have a helper method [..] for which I can call to grab a particular object rather than to remember which dependencies I need to hook up to get the object I require. \\
    My first question here is: should methods like these be tested? The only reason I can think of to test these methods would be to ensure that the correct dependencies are used and set up correctly. If the answer to the first question is yes, my second is: how? \cite{se_whichCodeStructure}}
\end{displayquote}
\vspace{.3em}

In the \textit{what} questions, we classified the discussions into seven categories. The \textit{What are the best practices to name test classes, methods, mocks, or paths}, \textit{What are the best practices for unit testing}, and \textit{What are the best practices for mocking} categories ask for coding standards, patterns, and guidelines for constructs in the test code. The \textit{What are the strategies and tools to find unused/dead code} and \textit{What are the tools to rename and organize tests} categories present tools for refactoring the test code to match the naming and structure of the production classes and improve the test code readability by removing unused code. The \textit{What are the mocking anti-patterns} and \textit{What are the test smells} categories ask for the anti-patterns and test smells definitions and catalogs that can occur in the test code. \looseness=-1

For example, we classified the following question in the \textit{What are the good practices for unit tests} category. With this question, the developer received guidance on which patterns to use while developing the test code, e.g., the \textit{Arrange, Act, Assert} (AAA) pattern for arranging and formatting the code. 

\vspace{.3em}
\begin{displayquote}
    \footnotesize
    \textsf{Usually, when talking about coding standards, we refer to the code of the program itself, but what about the unit tests? Are there certain coding standards guidelines that are unique to unit tests? \cite{se_bestPractices}} 
\end{displayquote} 
\vspace{.3em}

{\color{black}Conversely, we classified the following question \cite{se_testSmells} in the \textit{What are the test smells} category. The answers to this question provided more than 31 test code anti-patterns. }

\vspace{.3em}
\begin{displayquote}
    \footnotesize
    \textsf{There must be at least two key elements present to formally distinguish an actual anti-pattern from a simple bad habit, bad practice, or bad idea [...] Vote for the TDD anti-pattern that you have seen ``in the wild'' one time too many. \cite{se_testSmells}}
\end{displayquote}

\vspace{.3em}
\begin{center}
\begin{tcolorbox}[colback=azulbb, width=\linewidth] 
\textbf{Finding 1:} Developers are interested  in the practical experience of other developers to understand  test smells and decide whether and how to refactor the test code to fix them. 
\end{tcolorbox}
\end{center}
\vspace{.3em}

\subsection{Test code problems (RQ2)}
\label{sec:results-test-problems}

In RQ$_2$, we investigated the test smells discussed by developers in the Stack Exchange network. We followed Garousi et al.'s catalog \cite{Garousi2018_SurveySmells} to classify the posts.
~Although most 
question topics did not explicitly ask about a test smell, we could establish a link between the description of the test code problem and the test smell categorization. Therefore, we labeled the discussions without explicit test smells as \textbf{General Discussion (GD)} and the discussions with explicit test smell as \textbf{Specific Discussion (SD)} on test smells. We analyzed 36 SD with 46 test smells and 64 GD with 125 test smells.

\begin{figure*}[htbp]
\vspace{-0.5cm}
{\includegraphics[width=1.0\linewidth]{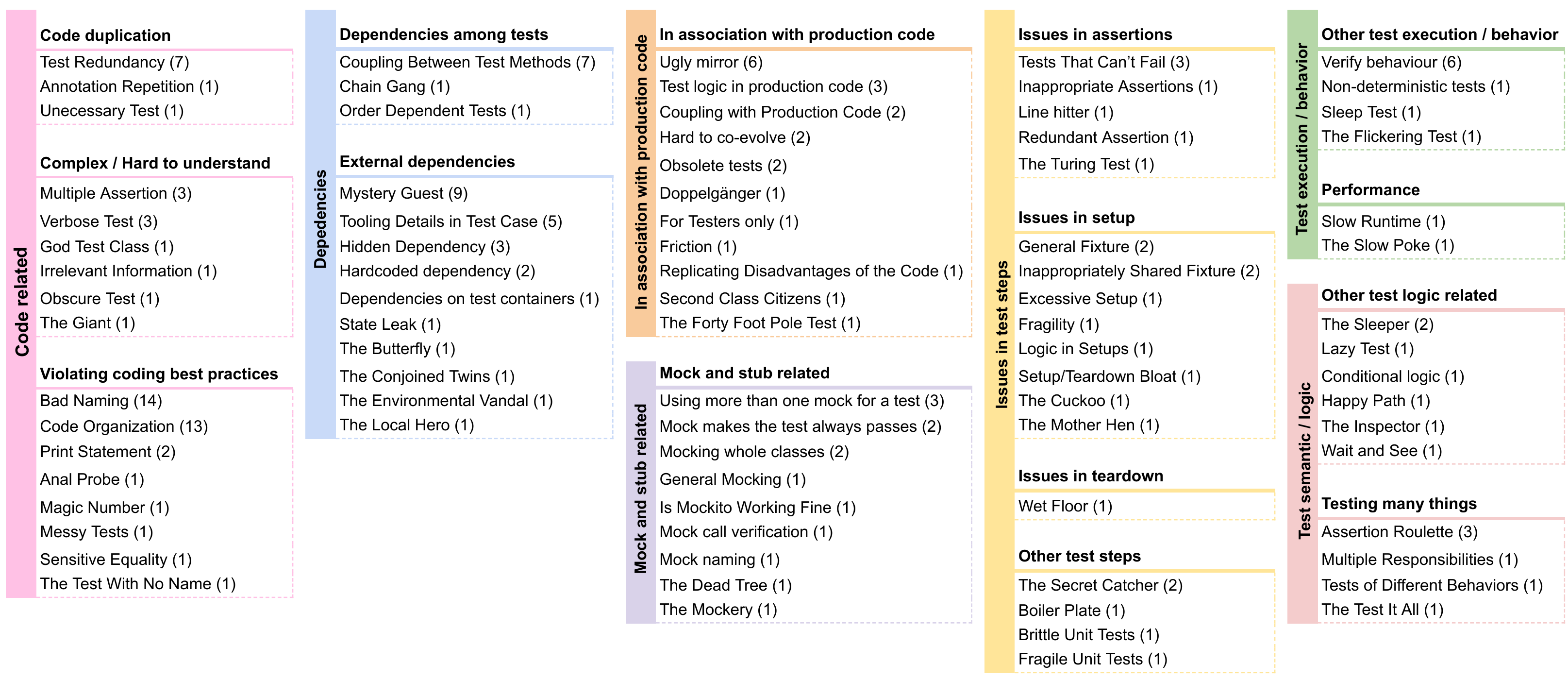}}
\caption{Classification of GD and SD on test smells according to Garousi's catalog of test smells \cite{Garousi2018_SurveySmells}.}
\vspace{-0.5cm}
\label{fig:test-problems}
\end{figure*}

Fig. \ref{fig:test-problems} presents the categorization of discussions into test smells. We found test smells composing seven of the eight top categories proposed by Garousi et al. \cite{Garousi2018_SurveySmells}. The \textit{Code related} category refers to test smells related to the test code duplication, long, complex, and hard-to-understand tests, and tests that do not follow coding best practices regarding naming conventions and code organization. The \textit{Dependencies} category refers to test smells related to dependencies within the test code or with external resources. The \textit{In association with production code} category refers to test smells related to coupling and dependencies between test and production code, making the tests hard to evolve. The \textit{Mock and stub related} category refers to test smells related to misusing mock objects and mocking verification. The \textit{Issues in test steps} category refers to occurring test smells in specific language constructs such as assertions and setup methods. The \textit{Test execution/behavior} category refers to test smells that can lead to unexpected results as non-determinism. The \textit{Test semantic/logic} category refers to test smells related to test logic and several responsibilities per test. The only category we did not find was the \textit{Design related} category, which mainly presents test smells related to page-object patterns in Selenium tests.

The seven categories from Fig. \ref{fig:test-problems} group 54 test smells. The \textit{Issues in test steps} category is the most diverse regarding test smells, comprising 18 test smells. Although the category is diverse, most of its test smells were discussed only once. Conversely, the \textit{Code related} category is diverse and comprises two of the most recurrent test smells in the discussions. The \textit{Bad Naming} test smell was the most frequent, with 14 occurrences. It describes the non-compliance with naming conventions for test code structures such as variables, methods, and classes. The \textit{Code Organization} test smell was the second most frequent, with 13 occurrences. It describes the non-compliance with test code organization as following the same production and test classes package hierarchy.

In addition, the \textit{Code related} category has the most test smells gathered from specific topics. The developers explicitly discussed ten out of 17 test smells of this category. Differently, we gathered the most test smells of other categories from the GD. While the \textit{In association with production code}, \textit{Mock and stub related} only contain GD, the \textit{Dependencies} and \textit{Issues in test steps} categories have three SD each, and the \textit{Test execution/behavior} and \textit{Test semantic/logic} categories have one SD each. It can indicate that developers face problems in different categories of test smells. Still, they know more about naming the test smells in the \textit{Code related} category.

\begin{center}
\begin{tcolorbox}[colback=azulbb, width=\linewidth] 
\textbf{Finding 2:} Developers usually ask whether something is a test smell or an anti-pattern, rather than referring to a particular one.
\end{tcolorbox}
\end{center}

\subsection{Test code refactorings (RQ3)}
\label{sec:results-test-refactorings}

In RQ$_3$, we investigated the solutions for handling test smells that developers suggest in the Stack Exchange network. We analyzed the actions suggested in the top answers of each discussion, resulting in 33 answers in the \textit{Fix} category, seven answers in the \textit{Capture} category, 70 answers in the \textit{Explain} category, and one answer in the \textit{Ignore} category.

In addition, we extracted the code-based answers, tools, and documents suggested in the answer topics to support developers in fixing test smells. The answers categorized into the \textit{Fix} category suggested 40 test code refactorings for fixing test smells. Less than half of the answers (13; 39.4\%) in this category presented code-based refactoring recommendations, i.e., the answers included code samples. In comparison, 32 answers (41.5\%) categorized into the \textit{Explain} and \textit{Capture} categories presented 64 patterns for organizing the test code and good practices for preventing test smells. Some answers (10; 12.9\%) presented examples of using patterns and good practices. The only answer in the \textit{Ignore} category discussed the trade-offs of refactoring a test method with many assertions to fix a test smell or keeping the assertions as documentation.

\begin{figure}[htbp]
{\includegraphics[width=\linewidth]{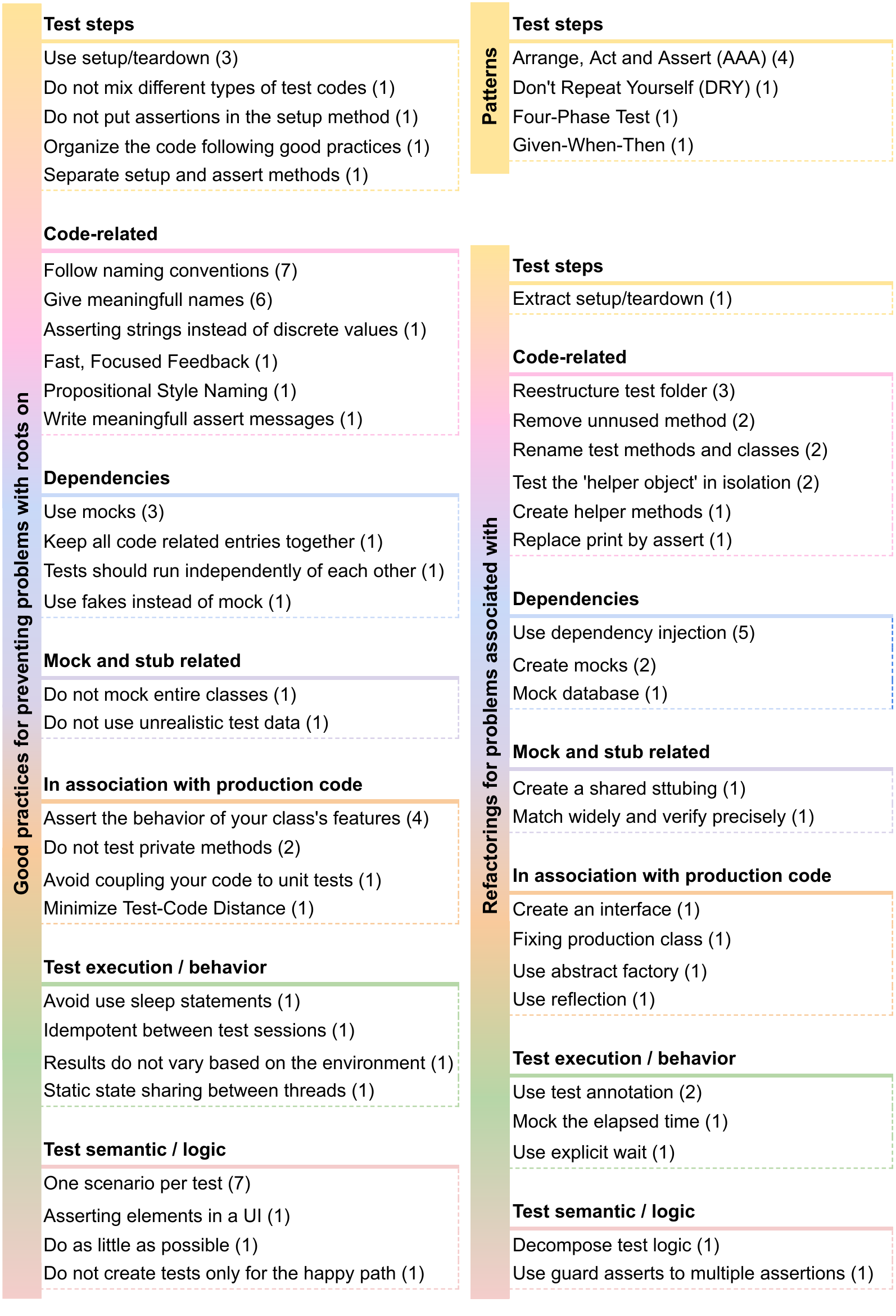}}
\caption{Suggestions of patterns for code organization, good practices, and test code refactorings for preventing and fixing test smells.}
\label{fig:test-refactorings}
\vspace{-0.5cm}
\end{figure}

Next, we analyzed the solutions proposed in the answers to remove duplicates and group similar answers according to the developers' definitions. Fig. \ref{fig:test-refactorings} presents 54 solutions classified into three categories: i) \textit{Good practices for preventing problems with roots on}, ii) \textit{Patterns}, and iii) \textit{Refactorings for problems associated with}. The colors represent the categories of test code problems from Fig. \ref{fig:test-problems}. The patterns, good practices, and test code refactorings aim to solve them.

The \textit{Patterns} category refers to patterns to structure the test code, avoiding issues in the test steps. Most solutions suggested organizing the test methods according to the AAA pattern \cite{beck2003test, khorikov2020unit}. This pattern organizes the test methods into three steps: (1) setup inputs and targets, (2) act on the target behavior, and (3) assert expected outcomes. It is also called the \textit{Given-When-Then} pattern in the \textit{Behavior Driven Development (BDD)} process \cite{smart2014bdd, korhonen2020automated}. The \textit{Four-Phase Test} pattern adds one step into that pattern: (4) reset to its pre-setup state \cite{xUnitWeb}.

The \textit{Good practices for preventing problems with roots on} category refers to good practices covering the top test problems presented in Fig. \ref{fig:test-problems}. The \textit{Test Steps} category presents good practices for organizing the test code. The \textit{Code-related} category suggests practices to provide useful information while naming variables and test methods and classes to facilitate the identification of failures caught by the tests. The \textit{Dependencies} category suggests practices for handling dependencies among tests and with external resources. The \textit{Mock and stub related} category brings insights into the correct usage of the mocking frameworks. The \textit{In association with production code} category suggests separating the responsibilities to avoid coupling between test and production codes. The \textit{Test execution/behavior} category aids in developing tests for async tasks of the production code. The \textit{Test semantic/logic} category presents good practices to avoid developing naive test cases and cover as many paths as possible. 

Similarly, the \textit{Refactoring for problems associated with} category refers to test code refactorings to solve the seven top problems presented in Fig. \ref{fig:test-problems}. The \textit{Test steps} category presents only one refactoring to extract common arrange code in test methods into setup/teardown methods. The \textit{Code-related} category presents refactorings to deal with code duplication and bad naming. The \textit{Dependencies} category suggests using mocks or dependency injection to handle external dependencies. The \textit{Mock and stub related} category presents refactorings to remove duplication related to mocks and ensure well-designed asserts. The \textit{Association with the production code} category presents refactorings in the production class that can require adaptations in the test code. The \textit{Test execution/behavior} category presents refactorings to handle async tasks by using specific features of the testing or mocking frameworks. Lastly, the \textit{Test semantic/logic} category presents refactorings to decompose the logic in test methods, reducing their complexity.

To guide the adoption of testing patterns, good practices, and test code refactorings, the developers pointed out 32 resources in their answers. Most resources are blogs and books (5; 21.7\% each) that present test anti-patterns and refactoring strategies to fix them. In addition, the developers suggested mocking frameworks (7; 30.4\%) and pointed out the documentation for constructs of testing and mocking frameworks (4; 17.4\%). Some developers also suggested online courses and videos (1; 4.4\% each).

\begin{center}
\begin{tcolorbox}[colback=azulbb, width=\linewidth] 
\textbf{Finding 3:} Developers more often discuss good practices and test patterns than code-based refactoring recommendations.
\end{tcolorbox}
\end{center}

\subsection{Test behavior (RQ4)}
\label{sec:results-test-behavior}

In RQ$_4$, we analyzed whether developers are concerned with keeping the test code behavior after performing refactorings to fix test smells. From 101 discussions, only eight (7.9\%) addressed test code behavior. In most discussions, the developers are interested in evolving legacy codes, understanding how to perform the refactoring phase of TDD, or improving the test code quality. 

After reading Martin Fowler's book \cite{fowler2018refactoring}, the developer decided to refactor the test code to organize and remove redundant code. The developer linked refactoring production and test codes, asking how to test the test code refactorings \cite{reading_books}. As a strategy to keep test code behavior, the answer suggests:\looseness=-1

\vspace{.3em}
\begin{displayquote}
\footnotesize
    \textsf{[...] The trick with complex refactoring of test code is to be able to run the tests against the system under test and get the same results. [...] \cite{reading_books}}
\end{displayquote}
\vspace{.3em}

In another discussion, the developer brings up definitions of refactoring and how to ensure that refactorings do not break the code behavior. Following those definitions, the developer's interest lies in understanding how to refactor smelly test codes and whether creating meta-tests helps keep the test code behavior \cite{behaviour_smelly_test}. The answer suggests: 

\vspace{.3em}
\begin{displayquote}
\footnotesize
    \textsf{When modifying tests, keep the SUT (System Under Test) unchanged. Tests and production code keep each other in check, so varying one while keeping the other locked is safest. [...] \cite{behaviour_smelly_test}}
\end{displayquote}
\vspace{.3em}

\begin{center}
\begin{tcolorbox}[colback=azulbb, width=\linewidth] 
\textbf{Finding 4:} Discussions on how to keep the test code behavior suggest 1) locking the production code while modifying the tests and 2) checking the results of the tests before and after performing the refactoring. There is no suggestion of tools or strategies to make this process more trustworthy. 
\end{tcolorbox}
\end{center}
\vspace{.3em}
\section{Discussion}\label{sec:Discussion}

This section discusses and interprets the results of each RQ we addressed in this study. Besides, this section points out some existing gaps.

In RQ\textsubscript{1}, we raised the challenges developers face regarding test code problems and the solutions to cope with them. Developers have to overcome the barrier of (1) convincing management and development teams of the implications of test smells for the test code quality, (2) co-evolving the production and test codes, and (3) keeping themselves up-to-date about the new constructs of programming languages and testing frameworks, as they emerge. Although all the discussions occurred after the first catalogs of test smells \cite{Deursen2001_TestRefactoring, Meszaros2003_AutomationManifesto}, we found that developers commonly ask for others' perceptions and practical experience on test code problems and proven solutions. The discussions on the Stack Exchange network highlight a gap between industry and academia, indicating the importance of effectively disclosing literature findings.    

In RQ\textsubscript{2}, we classified 54 test smells into eight high-level categories of test code problems, based on \cite{Garousi2018_SurveySmells}. In addition, we publicized the test smells definitions through an online catalog \cite{paginaweb} to help developers understand, prevent, detect, and refactor test smells. Practitioners could use the catalog to educate themselves on test smells concepts. To evolve the catalog, we invited researchers and developers to submit pull requests to update the website with test smells and their definitions. 

In RQ\textsubscript{3}, we listed the solutions for test code problems and classified them into good practices, test patterns, and test code refactorings. Most solutions explain how to deal with the test code problem but do not demonstrate them. In addition, one good practice or test code refactoring could prevent or refactor more than one test smell. As we could not establish a link between the good practices and test code refactorings to the specific test smells, we linked the good practices and test code refactorings to the top test code problems they aim to solve. Hence, exploring good practices and test code refactorings to fix test smells is essential. Most solutions pointed to blogs, books, documentation, and testing frameworks. Despite the research community's efforts in developing tools for handling test smells and other problems in the test code, the solutions of the Stack Exchange network did not point to any of them. Besides proposing new tools, academic studies should evaluate their usefulness in real-world contexts. 

In RQ\textsubscript{4}, we observed developers are concerned with keeping the test code behavior after the refactoring. However, no well-defined strategies or tools to aid developers in refactoring test code. It can lead developers to skepticism, missing an opportunity to improve test code quality. Therefore, researchers could focus on providing simple oracle mechanisms for test refactorings (e.g., mutation testing), or automating the catalog of refactorings in a static analyzer.
\section{Threats to Validity}\label{sec:ThreatstoValidity}

\textit{Search process: } 
We filtered the discussions related to the test code refactorings and test smells by applying a search string to the tags of each post. As developers can use different tags other than the ones we considered in our search string, we may have missed some discussions in exchange for reducing the number of non-relevant discussions returned with the search string. In addition, we only analyzed the accepted and top-voted responses for each discussion to analyze only the relevant answers that provided a refactoring action or explanation for the problems raised on the Stack Exchange network.

\textit{Generalization of findings: }
There are several technology-based question-and-answer websites on the Stack Exchange network. For this study, our scope focused on the top 3 websites (\textit{Stack Overflow}, \textit{Software Engineering}, and \textit{Code Review}), encompassing various programming-related topics. In addition, we applied a search string to the discussions' tags to limit our analysis to test code refactorings performed with Java programming language and the JUnit testing framework.

\textit{Manual analysis bias: }
We performed a peer review process to mitigate bias during the selection and classification of the discussions. First, we selected a set of potential discussions, and three researchers classified them independently. We achieved an agreement level of 0.61, following Kappa statistics. Also, the selection and classification processes involved discussions aiming to solve any potential conflicts. 
 
\textit{Popularity metric:}
To measure the popularity of a question and an answer, we considered counts of the number of scores on the posts. In addition, we did not consider the period of the questions (e.g., when the developers posted the questions). In our analysis, we did not distinguish between questions based on time, so there are no new questions with low counts of the number of scores and views. 

 \section{Related Work}\label{sec:RelatedWork}

\subsection{Developers’ perception of test smells}

Bavota et al. \cite{Bavota2012_testSmellsImpacts} performed the first study investigating the empirical evidence of test smells in 18 software systems and the software developers’ perception of the test smells effects on the code quality. Later, the authors extended that study \cite{bavota2015test} by investigating 27 software systems and surveying 61 developers. The authors observed a high diffusion of test smells in such a study, which may lead to issues concerning the comprehensibility and maintainability of test suites and production code.

Similarly, Tufano et al. \cite{Tufano2016_NatureTestSmells} surveyed 19 developers to investigate whether they could recognize occurrences of test smells in software projects. The results indicated developers do not recognize test smells and rarely remove them from the test code. Similarly, Junior et al. \cite{Junior2020_Survey, Junior2021_TaleEmpirical} conducted empirical studies to unveil how consciously software developers insert test smells. The results indicated that experienced professionals introduce test smells during their daily programming tasks, even when using standard practices from their companies.

Spadini et al. \cite{Spadini2020_SeverityThresholds} argued developers only sometimes perceive test smells as problematic, given the lack of thresholds to interpret them. The authors defined thresholds for nine test smells and empirically evaluated the perception of 31 developers on the proposed thresholds. As a result, the authors indicate that participants' perceptions agree with previously predefined severity thresholds and that test smells impact maintenance in test suites. Bai et al. \cite{Bai2022_ARTestSmell} investigated the impact of test smells on test learning with 42 computer science students. Results indicated some test smells become less severe or do not occur with the evolution of the testing frameworks.

Instead of analyzing developers' opinions through surveys and interviews, our analysis of Stack Exchange discussions addresses the main challenges developers face in handling test smells in practice. This work differs from state-of-the-art approaches by (i) addressing a larger set of developers through the Stack Exchange network discussions among many developers and (ii) studying the definitions of test smells in practice, providing additional context to earlier studies.

\subsection{Test code refactoring}

van Deursen et al. \cite{Deursen2001_TestRefactoring} introduced the concept of test smells and proposed a catalog describing test smells and refactorings to fix them. Complementary, Meszaros et al. \cite{Meszaros2003_AutomationManifesto} and Bowes et al. \cite{Bowes2017_HowGood} broadened the definition of test smells and listed relevant principles for test code. Later, Guerra et al. \cite{Guerra2007_Refactoring_Safely} explored those test smells definitions and proposed a catalog of 15 test code refactorings to fix them. In addition, the authors proposed a representation that can ease the analysis of whether the refactoring did not change the test code behavior. 

Turning the attention to the empirical studies, Kummer \cite{kummer2015categorising} studied whether 20 developers recognize and refactor test smells. Results pointed out developers refactor test smells by chance. Similarly, Soares et al. \cite{Soares2020_RefactoringTestSmells} investigated how developers refactor test code to eliminate test smells. The authors surveyed 73 open-source developers to assess their preference and motivation to choose between smelly and refactored test code samples. In another work, Soares et al. \cite{Soares2022_RefactoringJUnit5} investigated whether the JUnit 5 features help refactor test code to remove test smells. They conducted a mixed-method study to analyze the usage of the testing framework features in 485 Java open-source projects, identifying the features helpful for fixing test smells and proposing test code refactorings.

To understand how often and which strategies developers use to refactor the test code for fixing test smells, Santana et al. \cite{Santana2021_MultiMethod} surveyed 87 developers and interviewed eight other developers. Results indicated most participants consider relevant to refactor test smells but only sometimes do it. Similarly, Campos et al. \cite{Campos2021_DevelopersPerception} asked software developers to refactor the test code of their projects and remove existing test smells. The results indicated developers must learn how to refactor test code to remove the test smells.

The studies mentioned above asked for the developers’ opinions, while other studies mined the projects’ commits history to understand test code refactorings. Peruma et al. \cite{Peruma2020_RefactoringAndroid} investigated the relationship between refactorings and their effect on test smells in 250 open-source Android Apps. Results showed that refactoring operations in test and non-test files differ, and the refactorings co-occur with test smells. Kim et al. \cite{Kim2021_SecretLife} conducted an empirical study on the test smells evolution and maintenance in 12 open-source projects. The authors analyzed the commits that removed test smells and concluded the test smells removal was due to maintenance activities. Kim et al. \cite{Kim2021_TestAnnotation} studied the maintenance activities of developers on test annotations. They created a taxonomy by manually inspecting and classifying a sample of test annotation changes and documenting the motivations driving these changes. 

Differently, Alomar et al. \cite{Alomar2020_ExperienceRefactoring, Alomar2021_BehindScenes} discussed the importance of considering the developer's experience as part of solutions for code refactoring. The authors conducted an empirical study on 800 open-source projects to investigate the relationship between the developers' experience and the number of refactoring activities. As a result, the authors found several developers apply refactorings, but only a few are responsible for the production and test code. Furthermore, the authors reported no correlation between experience and motivation due to refactoring.

Instead of analyzing developers' opinions or refactoring activities through the projects' histories, our study analyzes discussions from the Stack Exchange network. Those discussions can anticipate the problems that developers face with their respective solutions in practice.

\subsection{Developers' perceptions of Stack Exchange topics}

Several studies have analyzed Stack Exchange network discussions on particular topics in the Software Engineering field \cite{Openja2020_ReleaseEngineering, Choi2015_CodeClones, Tian2019_ArchitectureSmells, Tahir2018_CodeSmells, Tahir2020_CodeSmells}. Openja et al. \cite{Openja2020_ReleaseEngineering} analyzed 260,023 release engineering questions using topic modeling. The authors examined the developers' topics of interest and their difficulties reported on Stack Overflow. The results indicated developers discussed 38 release engineering topics, among which software testing is the most challenging topic and the most important contributor to software quality assurance.

Other topics include discussions on smells that occur in different software artifacts. Choi et al. \cite{Choi2015_CodeClones} performed a preliminary study on 925 discussions on Stack Overflow about code clones. Results showed most discussions are related to refactoring with the need for more support for clone refactoring tools. Tian et al. \cite{Tian2019_ArchitectureSmells} analyzed the developers' perception of architectural smells through 207 discussions on Stack Overflow. The results indicated developers use general terms to describe architectural smells caused by violating architectural patterns and design principles.

Tahir et al. \cite{Tahir2018_CodeSmells} mined 17,126 Stack Overflow posts and manually analyzed the top 100. The results showed developers use Stack Overflow to ask for general code smell assessments rather than particular refactoring solutions. Tahir et al. \cite{Tahir2020_CodeSmells} also analyzed how developers discuss code smells and anti-patterns across three technical Stack Exchange sites. Results showed developers often discuss the downsides of implementing specific design patterns and flag them as potential anti-patterns that developers should avoid.

Regarding code refactoring to improve its overall design, Pinto and Kamei \cite{Pinto2013_RefactoringTools} mined Stack Overflow posts to study discussions around refactoring tools. The results indicated developers prefer multi-language refactoring tools. However, they do not use them due to usability issues and a lack of trust in the refactoring process. Peruma et al. \cite{Peruma2022_HowDoIRefactorThis} analyzed 9489 refactoring discussions on Stack Overflow to investigate the trends and challenges that the developers face in refactoring software artifacts in practice. The authors automatically assigned the discussions to one of the topics: (i) code optimization, (ii) architecture and design patterns, (iii) unit testing, (iv) tools and IDEs, and (v) database. As for unit testing, developers face challenges with writing test cases, mainly to accommodate refactored production code. As a result, the authors summarized the key challenges and conclusions for relevant stakeholders considering each topic.

In contrast, our study is the first to look into test smells within the discussions of the Stack Exchange sites. More specifically, our study investigates the challenges developers face in test code refactoring, the test code problems in practice, and the test code refactorings suggested to fix test smells.

\section{Conclusions and Future Work}\label{sec:conclusion}

In this study, we investigated the discussions about test smells and test code refactoring on the  Stack Exchange network. We aimed to leverage knowledge about the corrective actions developers take to deal with them and the main challenges developers face to correct test smells.

To accomplish our goal, we sorted the discussion topics into 30 categories. Each category describes questions developers ask when they face any issue during test code evolution. The yielded results indicate most topics lie in the \textit{Understanding which code structures to test (12)} and \textit{Understanding test smells (12)} categories. We observed developers are interested in others' perceptions and hands-on experience handling issues of test code. In addition, there is an indication the developers often ask whether test smells or anti-patterns are either good or bad testing practices than code-based refactoring recommendations. 

In future work, we plan to design a catalog of test smells and their definitions to help bridge the gap between tool supporters and developers to ensure that refactoring does not break tests.

\section*{Acknowledgments}
This work is partially supported by INES (www.ines.org.br), CNPq grant 465614/2014-0, FACEPE grants APQ-0399-1.03/17 and APQ/0388-1.03/14, CAPES grant 88887.136410/2017-00; and Coordenação de Aperfeiçoamento de Pessoal de Nível Superior - Brasil (CAPES) - Finance Code 001; and FAPESB grant BOL0599/2019 and BOL0188/2020.

\bibliographystyle{IEEEtran}
\bibliography{references}

\end{document}